\documentclass[10pt,conference]{IEEEtran}
\usepackage{cite}
\usepackage{amsmath,amssymb,amsfonts}
\usepackage{algorithmic}
\usepackage{textcomp}
\usepackage{xcolor}
\usepackage{todonotes}
\usepackage{comment}
\usepackage[listings,skins,theorems,breakable,most]{tcolorbox} 
\def\BibTeX{{\rm B\kern-.05em{\sc i\kern-.025em b}\kern-.08em
    T\kern-.1667em\lower.7ex\hbox{E}\kern-.125emX}}
\usepackage{booktabs} 
\usepackage{tabularx} 
\usepackage{balance} 
\usepackage{multirow} 
\usepackage{siunitx} 
\usepackage{hyperref}
\usepackage{graphicx,cleveref}
\newcommand{\multicell}[2][t]{\begin{tabular}[#1]{@{}l@{}}#2\end{tabular}} 
\newcommand{\ra}[1]{\renewcommand{\arraystretch}{#1}} 

\newcommand{\surveyquestion}[2]{
    \begin{center}
    \fbox{
        \parbox{0.8\columnwidth}{\textbf{Q#1:} #2
        }
    }
    \end{center}
}

\newcommand{\boxbox}[2]{
    \begin{center}
    \fbox{
        \parbox{#1\columnwidth}{#2
        }
    }
    \end{center}
}

\IEEEoverridecommandlockouts
\usepackage{tikz}
\newcommand\copyrighttext{%
  \footnotesize Copyright \textcopyright 2019 IEEE. Personal use of this material is permitted.
  Permission from IEEE must be obtained for all other uses, in any current or future 
  media, including reprinting/republishing this material for advertising or promotional 
  purposes, creating new collective works, for resale or redistribution to servers or 
  lists, or reuse of any copyrighted component of this work in other works. 
  
  \smallskip
  
  The definitive Version of Record was published in 2019 in Proceedings of the 41st International Conference on Software Engineering: Software Engineering Education and Training (ICSE-SEET'19). IEEE Press, 158--169. DOI: \texttt{\href{https://doi.org/10.1109/ICSE-SEET.2019.00025}{10.1109/ICSE-SEET.2019.00025}}
  }
\newcommand\copyrightnotice{%
\begin{tikzpicture}[remember picture,overlay]
\node[anchor=south,yshift=10pt] at (current page.south) {\fbox{\parbox{\dimexpr\textwidth-\fboxsep-\fboxrule\relax}{\copyrighttext}}};
\end{tikzpicture}%
}

\begin{document}

\title{Attitudes, Beliefs, and Development Data Concerning Agile Software Development Practices}

\author{
\IEEEauthorblockN{Christoph Matthies, Johannes Huegle, Tobias D{\"u}rschmid\IEEEauthorrefmark{1}, and Ralf Teusner}
\IEEEauthorblockA{Hasso Plattner Institute, University of Potsdam, Germany\\
Email: \{first.last\}@hpi.de}
\IEEEauthorblockA{\IEEEauthorrefmark{1} Now at Carnegie Mellon University, Pittsburgh, USA, Email: duerschmid@cmu.edu}
}

\maketitle

\copyrightnotice

\begin{abstract}
The perceptions and attitudes of developers impact how software projects are run and which development practices are employed in development teams.
Recent agile methodologies have taken this into account, focusing on collaboration and shared team culture.
In this research, we investigate the perceptions of agile development practices and their usage in Scrum software development teams.
Although perceptions collected through surveys of 42 participating students did not evolve significantly over time, our analyses show that the Scrum role significantly impacted participants' views of employed development practices.
We find that using the version control system according to agile ideas was consistently rated most related to the values of the Agile Manifesto.
Furthermore, we investigate how common software development artifacts can be used to gain insights into team behavior and present the development data measurements we employed.
We show that we can reliably detect well-defined agile practices, such Test-Driven Development, in this data and that usage of these practices coincided with participants' self-assessments.
\end{abstract}


\section{Introduction}
\label{sec:introduction}
As Software Engineering is an activity conducted by humans, developers' perceptions, beliefs, and attitudes towards software engineering practices significantly impact the development process~\cite{weinberg1971psychology, Devanbu2016}
If the application of a particular idea or method is regarded as more promising compared to an alternative, software developers will prefer the first method to the second.
The human aspects of Software Engineering, the perceptions and beliefs of the people performing the work, have repeatedly been recognized as important throughout the history of the discipline~\cite{Lenberg2015a}.
They are influenced by personal experience as well as a multitude of external factors, such as recorded second-hand experiences, arguments, anecdotes, community values, or scholarship~\cite{Ajzen2001, Bogart2016API}.
In this context, these bodies of knowledge, i.e., collections of beliefs, values and best practices are referred to as development methodologies or processes.
While much research has focused on the technical aspects and benefits of these practices, little research has been conducted on the perceptions and attitudes of development team members towards them and their influence on adoption.

\subsection{Agile Development Methodologies}
In terms of modern software engineering, agile software development methodologies, such as Scrum, have highlighted the importance of people, collaboration and teamwork~\cite{schwaber1997scrum}.
Scrum, currently the most popular agile software development methodology employed in industry~\cite{stateofagile11}, has been described as a ``process framework`` that is designed to manage work on complex products~\cite{Schwaber2017}.
The authors of the Scrum Guide emphasize its adaptability, pointing out that Scrum is specifically not a technique or definitive method, but rather a framework in which various processes can be employed~\cite{Schwaber2017}.
Which concrete practices are selected then depends on a team's assessments and prior beliefs.
The agile manifesto~\cite{fowler2001agile} defines agile values, such as ``responding to change over following a plan'', which are designed to serve as a guiding light for practice selection and process implementation.
However, due to their relative vagueness, these values themselves are open to interpretation by team members.

\subsection{Perception vs. Empirical Evidence}
Software developers are influenced in their work by their prior experiences.
Their attitudes and feelings towards certain agile development practices stem mainly from applying these practices in software projects~\cite{Devanbu2016}.
It is these attitudes towards agile practice application and developers' perceptions of them, that we aim to study.
While human factors are integral to the software development process, the main goal remains to produce a product that serves the intended purpose.
To this end, a large variety of primary artifacts, such as executable code and documentation as well as secondary, supportive development artifacts, such as commits in a version control repository or user stories containing requirements, are created.
All the data produced by software engineers on a daily basis, as part of regular development activities, is empirical evidence of the way that work is performed~\cite{Kalliamvakou2016} and represents a ``gold-mine of actionable information''~\cite{Guo2016}.
Combining and contrasting the perceptions of developers with the empirical evidence gathered from project data can yield insights into development teams not possible by relying on only a single one of these aspects.

In work that partly inspired this research, Devanbu et al. conducted a case study at Microsoft on the beliefs of software developers concerning the quality effects of distributed development and the relationship of these beliefs to the existent project data~\cite{Devanbu2016}.
They found that developers' opinions did not necessarily correspond to evidence collected in their projects.
While respondents from a project under study had formed beliefs that were in agreement with the collected project evidence, respondents from a second team held beliefs that were \textit{in}consistent with the project's actual data.
The authors conclude that further and more in-depth studies of the interplay between belief and evidence in software practice are needed.

Due to the issues involved with recruiting professional developers for studies, such as having to offer competitive fees and the requirement of not interfering with normal work~\cite{Dieste2013}, we, as much of the related literature~\cite{Sjøberg2005}, rely on software engineering students for our study.
In the context of an undergraduate collaborative software engineering course employing agile methodologies, which has become standard practice in universities~\cite{Paasivaara2017}, we study the usage and application of agile practices by developers and other Scrum roles and the related perceptions of these practices during software development.

\subsection{Research Questions}
The following research questions related to the perceptions of agile development teams and their relationship to agile practices and development data guide our work:
\begin{enumerate}
    \renewcommand{\labelenumi}{\textbf{RQ\arabic{enumi}}}
    \item \label{r1} How do perceptions of agile software development practices change during a software engineering course?
    \item \label{r2} What agile practices are perceived by students to be most related to agile values?
    \item \label{r3} Does the role within a Scrum team influence the perception of agile practices? 
    \item \label{r4} Do students in teams agree with each other in their assessments of the executed process?
    \item \label{r5} How can software development artifacts be used to gain insights into student behavior and the application of agile practices?
    \item \label{r6} What is the relationship between perceptions of agile practices and the development data measurements based on these?
\end{enumerate}

The remainder of this paper is structured as follows:
Section~\ref{sec:studycontext} presents the software engineering course which served as the context of this study.
Section~\ref{sec:survey} describes and discusses the survey that was employed to collect student perceptions of agile practice use in their teams.
In particular, Section~\ref{sec:survey_detail} details the motivations and backgrounds of each survey item.
Similarly, Section~\ref{sec:devdataanalysis} discusses the development data measurements that were employed, with Section~\ref{sec:measurementdefinition} providing details on their construction.
Section~\ref{sec:related_work} presents related work while Section~\ref{sec:conclusion} summarizes our findings and concludes.

\section{Related Work}
\label{sec:related_work}
Previous work on the human aspects of software engineering has also focused on the attitudes of developers towards development practices and attempts at relating these to empirical data.
Most closely related is the case study of Devanbu, Zimmermann, and Bird of developers at Microsoft~\cite{Devanbu2016}.
The authors study the prior beliefs of participants towards the quality effects of distributed development and investigate the relationships of these beliefs to empirical project data, including development data from code repositories.
The authors conclude that beliefs vary between projects, but do not necessarily correspond with the measurements gathered in that project.
Kuutila et al. employed daily experience sampling on well-being measures in order to determine connections between the self-reported affective states of developers and development data such as the number of commits or chat messages~\cite{Kuutila2018}.
While the authors report significant relationships between questionnaire answers and development data measures, they also point out that some of these relationships went contrary to previous lab-experiments in software engineering, highlighting the importance of study contexts.
Furthermore, there is a body of related research on the aspects that drive process methodology and application.
Hardgrave et al.~\cite{Hardgrave2003} report, based on a field study, that developers' opinions towards methodologies and their application are directly influenced by their perceptions of usefulness, social pressure, and compatibility.
Their findings suggest that organizational mandates related to methodology adoption are not sufficient to assure use of the methodology in a sustained manner.

\section{Study Context}
\label{sec:studycontext}
In this paper, we describe a case study on the application of agile practices and the perceptions of these practices by students of an undergraduate software engineering course.
The simulated real-world setting which is employed during the accompanying development project is ideally suited for collecting data on these aspects.
The main focus of the course is teaching collaboration and agile development best practices in a setting featuring multiple self-organizing student teams~\cite{Matthies2016c} jointly working in a real-world, hands-on project.
The course was most recently taught in the winter semester of 2017/18.
Students employ the same development tools that are used in industry, including issue trackers and version control systems, which are hosted on GitHub.
These systems, while allowing for easier collaboration, also contain a large amount of information on how students work in their teams~\cite{Rosen:2015:FSE, Matthiesb}.
In addition to students gaining experience with a popular industry tool, research has shown that students benefited from GitHub's transparent workflow~\cite{Feliciano2016}.
Usage of these tools predicted greater positive learning outcomes~\cite{Hsing2018}.
Due to the nature of the capstone course, participants already have a working knowledge of software development and have worked in small development teams as part of their undergraduate studies.
Nevertheless, project work is supported by junior research assistants acting as tutors who are present during regular, organized Scrum team meetings.
Regular lectures on agile development topics supplement the project work.

In order to gain insights into the perceptions regarding agile practices as well as the application of these by students, we employed a two-fold approach.
We conducted a survey on the implementation of agile practices at the end of every development iteration and we analyzed development artifacts for every student and their team in that same time frame.

\subsection{Development Process}
During the project, four iterations of the Scrum method are employed, which due to its more descriptive nature is especially suited for introducing agile concepts~\cite{Mahnic2015a}.
Following this, after students have gained some experience in their teams, the less descriptive, more dynamic Kanban method is introduced towards the end of the course.

\begin{figure}[htb]
    \centering
    \includegraphics[width=0.9\columnwidth]{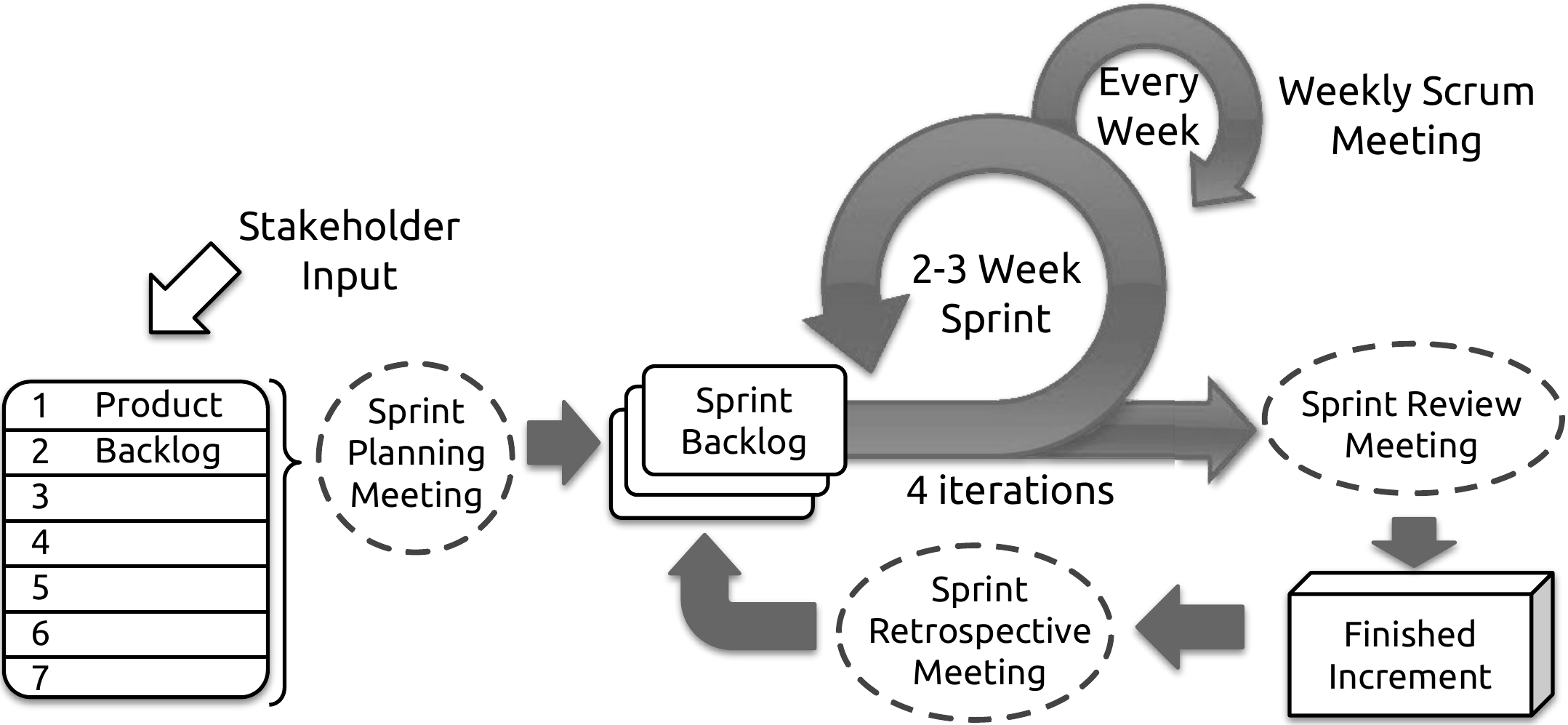}
    \caption{Overview of the adapted Scrum process employed in the course.}
    \label{fig:process}
\end{figure}

During the Scrum sprints of the development project, an adapted version of the Scrum process, which considers the time constraints of students, i.e., that other courses are running in parallel, is employed, see Figure~\ref{fig:process}.
In this paper, we focus on these four Scrum sprints.
Later iterations, which feature different development methodologies for shorter amounts of time are excluded.
The course relies heavily on the self-organization of agile teams~\cite{Hoda2010}.
Students form teams of up to 8 members and assign the roles of Product Owner (PO) and Scrum Master (SM) to team members, while all others act as developers.
The usual Scrum meetings of planning, sprint review, retrospective as well as a weekly stand-up synchronization meeting are held by every team in the presence of a tutor, who is able to provide consultation and instruction.
These meetings, through the use of surveys, as well as the development data produced by students during their unsupervised work time, serve as data collection opportunities to gain insights into team behavior.

\section{Survey}
\label{sec:survey}

\subsection{Construction}
The survey employed during the course started from a series of common agile practices and violations thereof drawn from software engineering research.
This related research deals with prevalent agile practices, such as Test-Driven Development and Collective Code Ownership~\cite{zazworka2010developers}, team meeting and collaboration best practices~\cite{Sletholt2012}, team environment and capability~\cite{chow2008survey} as well as practices that are common pitfalls for students~\cite{Matthies2016}.
In following the survey design of related literature~\cite{Devanbu2016}, we chose a small number of practices to focus on.
We used the following three criteria for selection:
(i) The topic is actively discussed during the course and participants have had experience with the practice or concept, allowing them to form an informed opinion.
(ii) Following or straying from practices is consequential for software projects.
(iii) The practice or concept allows gathering evidence based on already existent software development data of teams.

All claims of the survey are listed in Table~\ref{table:survey}.
We asked course participants to rate these on a 5-point Likert scale (strongly disagree, disagree, neutral, agree, strongly agree).

\begin{table}[htb]
    \caption{Overview of survey claims.}
    \ra{1.2}
    \begin{tabularx}{\columnwidth}{lX}
        \toprule
        \textbf{\#} & \textbf{Claim} \\
        \midrule
        1           & I wrote code using a test-driven approach \\
        2           & I practiced collective code ownership \\
        3           & The user stories of the sprint were too large \\
        4           & There were duplicates of user stories \\
        5           & I started implementing only shortly before the deadline \\
        6           & We followed the ``check in early, check in often'' principle \\
        7           & I worked on too many user stories simultaneously \\
        8           & We conducted useful code reviews \\
        9           & Our team has successfully implemented the agile values \\
        \bottomrule
    \end{tabularx}
    \label{table:survey}
\end{table}

\subsection{Administering the Survey}
The survey was presented to the students during the sprint review meeting at the end of each sprint, see Figure~\ref{fig:process}.
The questionnaire was given to participants as a hard copy, to be filled out by hand, before or after the meeting, according to the time of arrival.
The survey indicated that the information given by the students had no influence on their grading and that results will be published in anonymized form.
Furthermore, participants were informed beforehand, that survey results were used solely for research purposes.
Students were encouraged to answer only the questions they were able to answer, e.g., to ignore programming questions if they did not program during the development iteration or to ignore questions that were not relevant to their Scrum role.

\subsection{Survey Claims in Detail}
\label{sec:survey_detail}

This section presents the claims of the conducted survey in detail and provides background information on the relevance of the development practice in question.

\subsubsection{Test-first Programming Approach}
Test-driven development (TDD) describes a very short, repeated, software development approach: test cases are written for the user story to be implemented, then code is written to pass the new tests~\cite{beck2003}.
Previous research has highlighted the value of TDD in education and has established positive correlations between the adherence to TDD and students' solution code quality and their thoroughness of testing~\cite{Buffardi2012}.
Sampath showed that introducing TDD at higher learning levels, such as the capstone course described here, held more promise than at lower levels~\cite{Sampath2014}.
Application of the TDD method has been shown to have a positive effect on the quality of software design and assures that code is always tested~\cite{Madeyski10}.

\surveyquestion{1}{I wrote code using a test-driven approach}

\subsubsection{Code Ownership}
Collective Code Ownership (CCO) is one of XP's core practices, focusing on the shared responsibility of teams~\cite{Beck2000}.
It describes the idea that no developer ``owns'' the code.
Instead, anyone on the team should improve any part of the software at any time if they see the opportunity to do so~\cite{Beck2000}.
CCO allows any developer to maintain code if another is busy~\cite{Fitzgerald2006} and enables all code to get the benefit of many developers' attention, increasing code quality~\cite{Lindstrom2004}.
It furthermore plays a role in encouraging team interaction, cohesiveness, and creativity~\cite{Nordberg03}.

\surveyquestion{2}{I practiced collective code ownership}

\subsubsection{User Story Size}
User Stories offer a natural language description of a software system's features that hold value to a user or purchaser~\cite{Cohn2004}.
Empirical research has shown a correlation between high-quality requirements and project success~\cite{Kamata2007}.
However, measuring the quality of user stories is an ongoing challenge and little research is currently available~\cite{Lucassen2015}.
In a survey with practitioners, Lucassen et al. found that the largest group of respondents (39.5\%) did not use user story quality guidelines at all, while 33\% indicated they followed self-defined guidelines~\cite{Lucassen2016a}.
Only 23.5\% of respondents reported the use of the popular INVEST list~\cite{wake2003invest} to gauge user story quality.
The ``S'' in this mnemonic stands for small, which should apply to the amount of time required to complete the story, but also to the textual description itself~\cite{wake2003invest}.

\surveyquestion{3}{The user stories of the sprint were too large}

\subsubsection{User Story Duplication}
The sprint backlog, filled with user stories, is the main artifact that tracks what a development team is working on in a sprint.
Mismanaging the product or sprint backlog by including highly similar stories can lead to duplicated effort and ``software development waste''~\cite{Sedano2017}.
Therefore, every user story should be unique and duplicates should be avoided~\cite{Lucassen2015}.
While all user stories should follow a similar layout and format, they should not be similar in content.

\surveyquestion{4}{There were duplicates of user stories}

\subsubsection{Work Distribution}
Agile projects should follow a work pace that can be ``sustained indefinitely''~\cite{fowler2001agile} and which does not put unduly stress on team members.
``Death marches''~\cite{Yourdon1999}, the practice of trying to force a development 
team to meet, possibly unrealistic, deadlines through long working days and last-minute fixes have been shown to not be productive and to not produce quality software.
As Lindstrom and Jeffries point out, development teams are ``in it to win, not to die''~\cite{Lindstrom2004}.
If work is not evenly distributed over the sprint and more work is performed towards the end this can make the development process unproductive.
Team meetings conducted throughout the sprint cannot discuss the work, as it is not ready yet, blockers and dependencies with other teams might not be communicated on time and code reviews are hindered~\cite{Matthies2016}.

\surveyquestion{5}{I started implementing only shortly before the deadline}

\subsubsection{Version Control Best Practices}
Part of the agile work ethos of maximizing productivity on a week by week and sprint by sprint basis~\cite{Lindstrom2004} is an efficient usage of version control systems.
A prevailing motto is to ``check in early, check in often''~\cite{atwood08}.
This idea can help to reduce or prevent long and error-prone merges~\cite{Mandala2013}.
Commits can be seen as snapshots of development activity and the current state of the project.
Frequent, small commits can, therefore, provide an essential tool to retrace and understand development efforts and can be reviewed by peers more easily.
Research has shown that small changesets are significantly less associated with risk and defects than larger changes~\cite{Purushothaman2005}.
We encourage student teams not to hide the proverbial ``sausage making'', i.e., not to disguise how a piece of software was developed and evolved over time by retroactively changing repository history~\cite{Robertson2018}.

\surveyquestion{6}{We followed the ``check in early, check in often'' principle}

\subsubsection{Working in Parallel}
Focusing on working on a single task, instead of trying to multitask, leads to increased developer focus~\cite{Middleton2012} and less context switching overhead~\cite{Johnson2003}.
The number of user stories developers should work on per day or sprint unit cannot be stated generally as it is highly dependent on context.
In Lean thinking, strongly related to agile methods, anything which does not add value to the customer is considered ``waste''~\cite{Sedano2017}.
This includes context switches and duplicated effort by developers, and work handoff delays.
Little's Law is a basic manufacturing principle, also applicable to software development, which describes the relationship between the cycle time of a production system as the ratio of work in progress (WIP) and system throughput~\cite{Little2011}.
Cycle time is the duration between the start of work on user story and the time of it going live.
This time span can be reduced by either reducing the number of WIP user stories or by increasing throughput.
As throughput increases are usually limited by the number of team members, the cycle time can only reasonably be decreased by reducing the maximum number of user stories being worked on.
Delivering stories faster allows for quicker feedback from customers and other stakeholders.
Furthermore, limiting the work in progress in a team can help in identifying opportunities for process improvement~\cite{Middleton2012}.
When loaded down with too much WIP, available bandwidth to observe and analyze the executed process is diminished.
\surveyquestion{7}{I worked on too many user stories simultaneously}

\subsubsection{Code Reviews}
Code reviews are examinations of source code in order to improve the quality of software.
Reviewing code as part of pull requests in GitHub, following a less formal and more lightweight approach, has been called ``modern code review'', in contrast to the previously employed formal code inspections~\cite{Bacchelli2013}.
Code reviews by peers routinely catch around 60\% of defects~\cite{Boehm2001}.
However, finding defects is not the only useful outcome of performing modern code reviews.
They can also lead to code improvements unrelated to defects, e.g., unifying coding styles, and can aid in knowledge transfer.
Furthermore, code reviews have been connected to increased team awareness, the creation of alternative problem solutions as well as code and change understanding~\cite{Bacchelli2013}.
\surveyquestion{8}{We conducted useful code reviews}

\subsubsection{Agile Values}
The Agile Manifesto is the foundation of agile methods.
It describes the principles and values that the signatories believe characterize these methods.
These include principles such as early and continuous delivery of valuable software, welcoming changing requirements and building projects around motivated individuals~\cite{fowler2001agile}.
Scrum defines five the values of commitment, focus, openness, respect, and courage~\cite{Kropp2016a} that agile teams should follow in order to facilitate collaboration.
Extreme Programming similarly defines values that a team should focus on to guide development.
They include simplicity, communication, feedback, respect, and courage~\cite{Beck2000}.
Kropp et al. consider these values, compared to technical practices and collaboration practices, to be hardest to teach and learn as they require students to adapt their attitudes and evaluations~\cite{Kropp2016a}.
\surveyquestion{9}{Our team has successfully implemented agile values}

\subsection{Survey Results}
\num{42} students (\num{3} female and \num{39} male), participated in the course.
The students formed six teams: One of six students, one of eight students, and four of seven students.
Although participation in the survey was voluntary, every student who attended their team's Scrum meetings filled in a questionnaire.

\begin{figure}[htbp]
    \centering
	\includegraphics[scale=0.6]{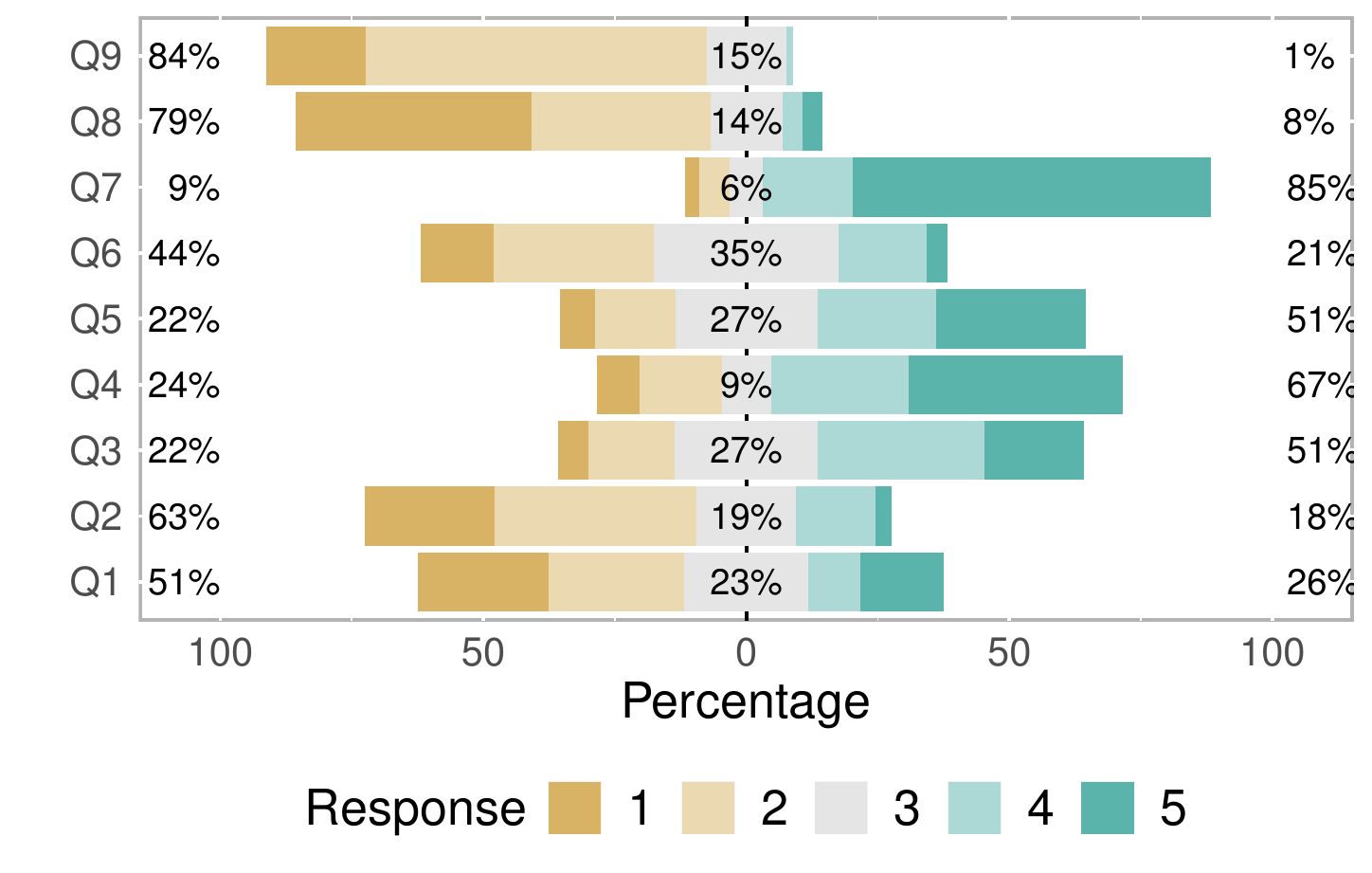}\vspace{-10pt}
	\caption{Summary of responses to survey questions Q1-Q9 on the 5-point Likert scale (1 ''strongly agree´´ to 5 ''strongly disagree´´) over teams and sprints.}
	\label{fig:OverallBarplots}
\end{figure}

\begin{table}[htb]
    \centering
    \caption{Descriptive statistics of responses to survey questions \newline
    Q1 - Q9 on a 5-point Likert scale \newline
    (from 1 ''strongly agree´´ to 5 ''strongly disagree´´).}
    \begin{tabular}{@{}lrrrrrrrrr@{}}
    	\toprule
    	                       &   Q1 &   Q2 &    Q3 &    Q4 &    Q5 &   Q6 &   Q7 &   Q8 &   Q9 \\
    	\midrule
    	Valid                  &  132 &  138 &   158 &   160 &   137 &  131 &  141 &  132 &  159 \\
    	Missing                &   36 &   30 &    10 &     8 &    31 &   37 &   27 &   36 &    9 \\
    	Mean                   &  2.7 &  2.3 &   3.4 &   3.8 &   3.5 &  2.7 &  4.4 &  1.9 &  2.0 \\
    	Median                 &  2.0 &  2.0 &   4.0 &   4.0 &   4.0 &  3.0 &  5.0 &  2.0 &  2.0 \\
    	Stdev                  &  1.4 &  1.1 &   1.1 &   1.3 &   1.2 &  1.0 &  1.0 &  1.0 &  0.6 \\
    	Skewness               &  0.4 &  0.6 &  -0.3 &  -0.8 &  -0.3 &  0.2 & -1.9 &  1.3 &  0.3 \\
    	StderrSkew             &  0.2 &  0.2 &   0.2 &   0.2 &   0.2 &  0.2 &  0.2 &  0.2 &  0.2 \\
    	\bottomrule
    \end{tabular}
    \label{table:survey_stats}
\end{table}
 Table~\ref{table:survey_stats} presents the mean and median ratings as well as the standard deviations of all the claims in the survey. Figure~\ref{fig:OverallBarplots} displays summarized participant answers using centered bar plots.
High agreement with a claim was coded as 1, disagreement as 5 on the Likert scale.
Therefore, the lower the mean, the more survey participants agreed with the stated claim. 
The higher the variance, the higher the disagreement was within respondents regarding the claim.
Over all sprints, the claims that survey participants disagreed with most concerned usage of user stories.

Participants stated that they on average had not worked on multiple user stories simultaneously during an iteration (Q7, mean of 4.4) and that there were few issues with duplicate user stories (Q4, mean of 3.8).
The claim that students on average agreed with the most concerned conducting useful code reviews (Q8, mean of 1.9), closely followed by high agreement on the claim that the team had successfully implemented agile values in their project work (Q9, average 2.0).

\subsection{Survey Discussion}
As the collected Likert data is ordinal we have to build our statistical inference on nonparametric methods~\cite{hollander2013nonparametric}.

\subsubsection{Perception change during the course (RQ\ref{r1})}
Our survey design involves repeated measurements, therefore, the first research question examines changes of developers' perceptions over in these measurements over time. 

\begin{figure}[ht]
\begin{center}
	\includegraphics[width=\columnwidth]{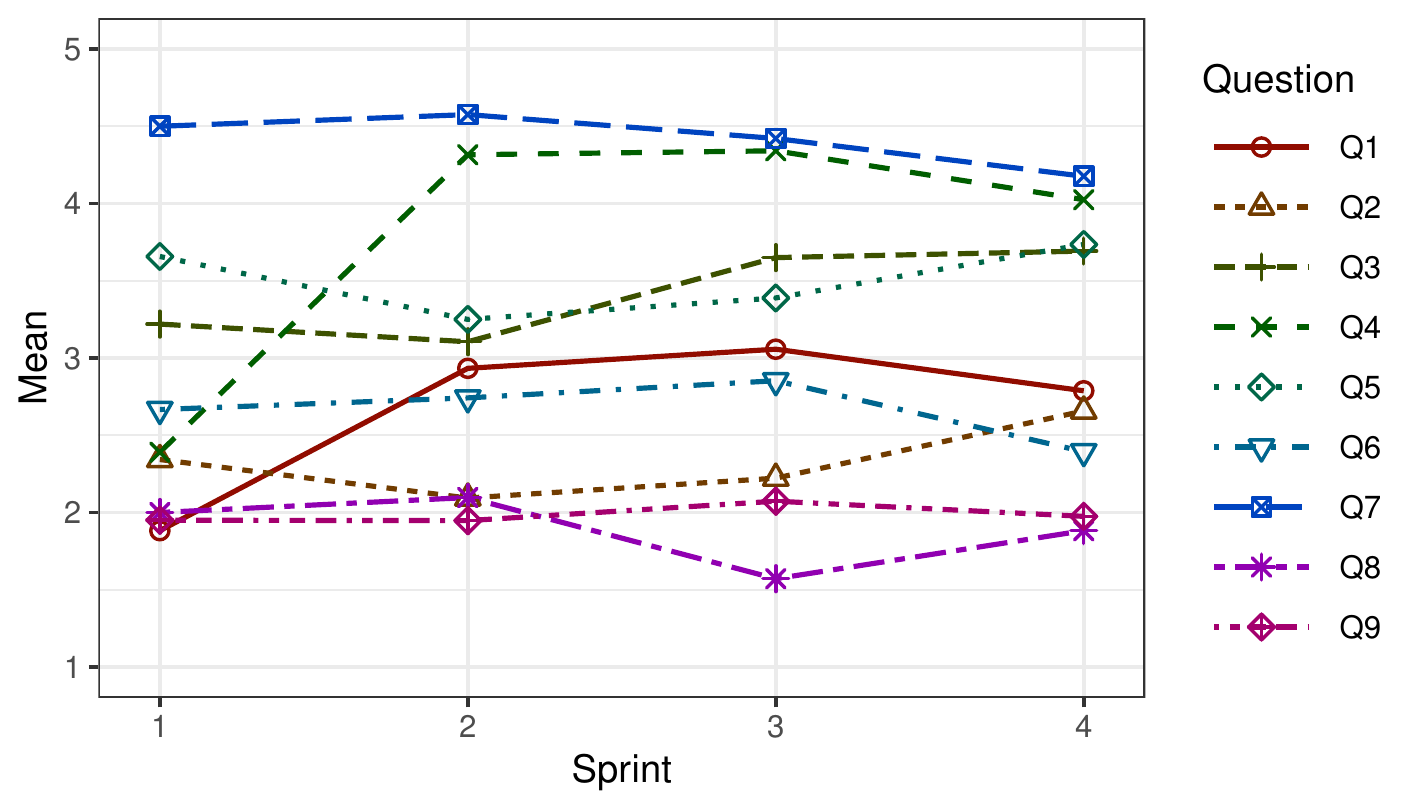}
\end{center}\vspace{-15pt}
	\caption{Means of survey questions (Q1-Q9) on the 5-point Likert scale (1 ''strongly agree´´ to 5 ''strongly disagree´´) and their developments over the sprints (1-4)}
	\label{fig:MeansPlot}
\end{figure}

Figure~\ref{fig:MeansPlot} depicts the history of survey response means over the four sprints, giving an overview of the variability of participants' responses.
Besides the responses to survey claim Q4, the developments of the means are comparatively stable indicating that the perceptions of participants do not change.
This intuition is supported by the application of a Friedman test, which allows analyzing effects on Likert responses from a repeated measurement experimental design~\cite{hollander2013nonparametric}.

\begin{table}[htbp]
    \caption{Friedman rank sum $\chi^2$, and statistical significance on Q1 - Q9 for changes in responses during the course}
    \label{table:friedman}
    \begin{minipage}{.4\columnwidth}
      \centering
        \begin{tabular}{@{}crl@{}}
            \toprule
            Question & $\chi^2$ & $p$-value  \\
            \midrule
            Q1 & $7.340$ & $0.062$  \\
            Q2 & $5.281$ & $0.152$ \\
            Q3 & $6.211$ & $0.101$ \\
            Q4 & $34.658$ & $1.4e^{-7}$*** \\
            \bottomrule
            \multicolumn{3}{l}{* p$<$.05, ** p$<$.01, *** p$<$.001}
            \end{tabular}
    \end{minipage}%
    \begin{minipage}{.7\columnwidth}
      \centering
        \begin{tabular}{@{}crl@{}}
            \toprule
            Question & $\chi^2$ & $p$-value  \\
            \midrule
            Q5 & $3.969$ & $0.265$ \\
            Q6 & $1.980$ & $0.577$ \\
            Q7 & $4.686$ & $0.196$ \\
            Q8 & $6.423$ & $0.093$ \\
            Q9 & $1.6343$ & $0.652$ \\
            \bottomrule
            \end{tabular}
    \end{minipage} 
\end{table}

As depicted in Table~\ref{table:friedman}, the conducted Friedman tests reveal that only the responses to the claim of duplicated user stories (Q4) show a significant change over time ($p < .01$).
Moreover, a post hoc test using Wilcoxon signed rank test with Bonferroni correction showed the significant change during the second sprint as there are significant changes ($p < .01$) in the responses at the end of the first sprint compared to other responses.
This finding can be related to the course context.
At the beginning of the project, the team's Product Owners need to work together to create initial user stories for the first development sprint from interviews with the product customer.
While they are guided by extensive tutoring, our experience has shown that this task of collaboration in combination with new tasks is challenging for students~\cite{Matthies2016a}.
As a result, the distribution of responsibilities and user stories between teams can lead to duplications of user stories.
As development goes on, these duplications are detected and removed.
For the other survey claims, the findings of reluctance to change in development processes are in line with related work.
Zazworka et al. report that even with specific interventions in teams, agile practices continued to be violated.
The authors attribute this in part to the notion, that during project work, satisfying the customer had higher a priority than following the steps of a defined practice~\cite{zazworka2010developers}.

\boxbox{0.9}{\textbf{Research Question \ref{r1}}:
    How do perceptions of agile software development practices change during a software engineering course? \\
    \textbf{Answer:}
    Only change: perceived duplication of user stories significantly decreased after the initial sprint.
}

\subsubsection{Agile practices and agile values (RQ\ref{r2})}
When examining associations within survey responses, the relationships of perceptions of agile practice application (Q1-Q8) with the assessments of agile value implementation (Q9) are especially interesting.
They allow for an identification of those practices that survey participants most relate to adopting the mindset of agile methodologies.
In the context of Likert data, a well-known measure of association is Kendall's Tau whose statistical significance can be tested by the application of Fisher's exact test of independence~\cite{hollander2013nonparametric}.
The computed Kendall's Tau coefficients and $p$-values are depicted in Table~\ref{table:KendallQ9}.

\begin{table}[htbp]
    \centering
    \caption{Kendall's $\tau$ coefficients, Fisher's Z, and statistical significance as measure of relationship between agile practices and to agile values}
        \begin{tabular}{crrl}
            \toprule
            Relationship  & $\tau$ & Z & $p$-value  \\
            \midrule
            Q1 to Q9 & $0.049$ & $0.650$ & $0.517$  \\
            Q2 to Q9 & $0.149$ & $1.978$ & $0.048$* \\
            Q3 to Q9 & $-0.114$ & $-1.638$ & $0.102$ \\ 
            Q4 to Q9 & $-0.020$ & $-0.289$ & $0.773$ \\
            Q5 to Q9 & $-0.212$ & $-2.827$ & $0.005$** \\
            Q6 to Q9 & $0.238$ & $3.075$ & $0.002$** \\
            Q7 to Q9 & $-0.040$ & $-0.519$ & $0.606$ \\ 
            Q8 to Q9 & $0.084$ & $1.060$  & $0.291$ \\
            \bottomrule
            \multicolumn{4}{l}{* p $<$ .05, ** p $<$ .01, *** p $<$ .001}
        \end{tabular}
   \label{table:KendallQ9}
\end{table}

We find significant relationships between the ratings of successful agile value implementation (Q9) to survey claims concerning practicing Collective Code Ownership (Q2, $\tau=0.15$, $p<.05$), not starting implementation shortly before the deadline (Q5, $\tau=-0.21$, $p<.01$), and following the ``check in early, check in often'' principle (Q6, $\tau=0.24$, $p<.01$).

\boxbox{0.9}{\textbf{Research Question \ref{r2}}:
    What agile practices are perceived by students to be most related to agile values? \\
    \textbf{Answer:}
    There are significant relationships of perceived success in implementing agile values with Collective Code Ownership, checking in early and often and not working at the last minute.
}

\subsubsection{The influence of Scrum roles (RQ\ref{r3})}
A major principle of the Scrum method is the division of a team into roles.
These include a Product Owner, a Scrum Master, and the development team. 
Hence, the third research question examines the influence of the team members' Scrum role on their perception of agile practices.

\begin{figure}[h]
\begin{center}
	\includegraphics[scale=0.6]{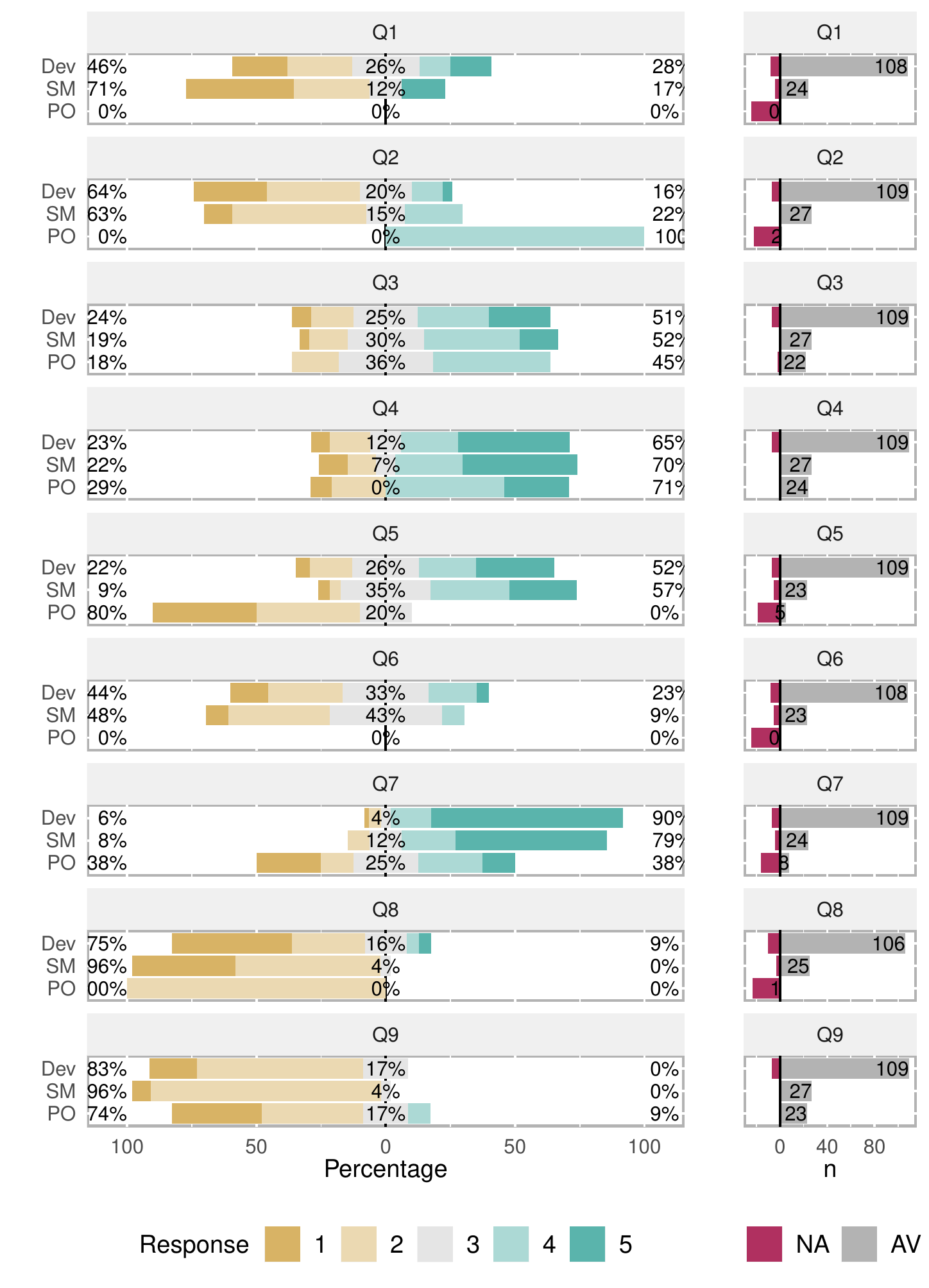}\vspace{-10pt}
	\caption{Ratings of survey claims on the 5 point Likert scale (1 ``strongly agree'' to 5 ``strongly disagree'') on the left side, and overview on missing data divided according to role on the right side}
	\label{fig:RoleBarplots}
\end{center}
\end{figure}

As depicted in Figure~\ref{fig:RoleBarplots} Product Owners did not rate the majority of survey claims regarding agile development practices, as they were mostly not directly involved with development tasks and implementation.
This aggravates the statistical examination for this role.
Furthermore, we found only small differences in the responses among roles.
Disregarding Product Owners, the bar plots indicate a small variation of responses to survey claims Q1, Q5, and Q7.

A nonparametric statistical method to compare the responses against between-subject factors, such as the roles of the survey attendees, is provided by the Kruskal-Wallis rank sum test~\cite{hollander2013nonparametric}.
As depicted in Table~\ref{table:kruskall}, the Kruskal-Wallis rank sum tests reveal significant effects of the attendees' role on the perception of TDD (Q1, $\chi^2=4.126$, $p < .05$), last-minute work (Q5, $\chi^2=8.671$, $p < .05$), and working on too many user stories simultaneously (Q7, $\chi^2=17.42$, $p < .001$).

\begin{table}[htbp]
    \centering
    \caption{Kruskal-Wallis rank sum $\chi^2$ and statistical significance of role effects on responses}
    \label{table:kruskall}
      \begin{minipage}{.4\columnwidth}
        \centering
        \begin{tabular}{@{}crl@{}}
            \toprule
            Question  & $\chi^2$ & $p$-value  \\
            \midrule
            Q1 & $4.126$ & $0.042$*  \\
            Q2 & $5.336$ & $0.069$ \\
            Q3 & $0.769$ & $0.681$ \\
            Q4 & $1.038$ & $0.595$ \\
            Q5 & $8.671$ & $0.014$* \\
            \bottomrule
            \end{tabular}
    \end{minipage}%
    \begin{minipage}{.7\columnwidth}
      \centering
        \begin{tabular}{@{}crl@{}}
            \toprule
            Question  & $\chi^2$ & $p$-value  \\
            \midrule
            Q6 & $0.451$ & $0.502$ \\
            Q7 & $17.42$ & $1.649 e^{-4}$*** \\
            Q8 & $0.754$ & $0.888$ \\
            Q9 & $0.239$ & $0.888$\\
            \bottomrule
            \multicolumn{3}{l}{* p$<$.05, ** p$<$.01, *** p$<$.001}
        \end{tabular}
    \end{minipage} 
\end{table}

In the context of TDD (Q1), a Dunn's test of multiple comparisons based on using rank sums with Bonferroni correction showed a significant difference in the perception between Scrum Masters compared to developers ($Z=2.03$, $p<.05$).
Here, the amount of missing values, as depicted in~\ref{fig:RoleBarplots}, precludes a statistical inference compared to Product Owners.

The same post hoc test showed significant differences in perceptions regarding last-minute work (Q5) of Product Owners, who worked on user stories, compared to both developers ($Z=2.859$, $p<.05$) and Scrum Masters ($Z=2.860$, $p<.05$), who were mainly concerned with coding activities.

Regarding perceptions of working on too many user stories simultaneously (Q7), there exists a significant difference between Product Owners and developers ($Z=4.036$, $p<.01$) as well as Scrum Masters ($Z=2.791$, $p<.05$).
These findings highlight the different nature of the Product Owner role, concerned with backlog and user story maintenance, and the other team roles who are performing the coding work.
As POs work almost exclusively with user stories, they have a higher chance of working simultaneously on them.
While the tasks of the Product Owner are mostly concentrated at the beginning of a user story's lifecycle, i.e., creating it, interacting with the customer and iteratively refining it, the work of implementing it by the developers must necessarily follow after this initial step.
As Product Owners are not actively involved with user story implementation~\cite{Schwaber2017}, they are mostly unaware of the (technical) challenges involved in the implementation.

\boxbox{0.9}{\textbf{Research Question \ref{r3}}:
    Does the role within a Scrum team influence the perception of agile practices?\\
    \textbf{Answer:}
    Scrum roles influenced the perceptions of few practices:
    SMs perceived TDD usage to be higher than developers. POs perceived they had started their management tasks only shortly before deadline more strongly than developers and SMs did regarding their development tasks.
}
    
\subsubsection{Agreement within teams (RQ\ref{r4})}
In Software Engineering the perception of agile practices and agile values is highly influenced by personal experience~\cite{Devanbu2016}.
Since all members of a team interact during the development process, the fourth research question examines whether this interaction leads to an agreement in the assessments within a team. In order to quantify the extent of agreement between the members of a team, we calculate Krippendorff's alpha that is a statistical measure of the inter-rater reliability~\cite{hayes2007answering}.

\begin{table}[h]
    \centering
    \caption{Krippendorff's alpha as measure of inter-rater agreement regarding the survey responses for the six development teams}
        \begin{tabular}{c|rrrrrr}
            \toprule
            Team & 1 & 2 &  3 & 4 & 5 & 6 \\
            \midrule
            $\alpha$ & $0.272$ & $0.424$ & $0.554$ & $0.608$ & $0.425$ & $0.285$ \\
            \bottomrule
        \end{tabular}
   \label{table:Krippendorff}
\end{table}

As $\alpha = 1.00$ denotes a perfect agreement and $\alpha = 0.00$ represents the total absence of agreement in a team, Table \ref{table:Krippendorff} reveals a moderate level of agreement in the six teams. In our survey, merely team 3 and team 4 show a substantial level of agreement in their perception of agile practices and agile values.
Moreover, team 1 and team 6 reveal a tendency towards disagreement.
Note that with moderate Krippendorff's alphas Scrum Masters ($\alpha = 0.437$) as well as developers ($\alpha = 0.408$) demonstrate a higher level of agreement across all teams while Product Owners show only a slight level of agreement ($\alpha = 0.238$).
This may be an indication that Product Owners, with their role requiring more focus on their teams' outcomes than the application of agile best practices, have a tendency towards more divergent perceptions.
A slight increase in the level of agreement in team 1 ($\alpha = 0.288$) and team 6 ($\alpha = 0.313$) when excluding the teams' product owners from the analysis supports this notion. 

\boxbox{0.9}{\textbf{Research Question \ref{r4}}:
    Do students in teams agree with each other in their assessments of the executed process?\\
    \textbf{Answer:}
    There are moderate levels of agreement in teams. Teams differ: two show a tendency towards disagreement, two share substantial agreement.
}

\section{Development Data Analysis}
\label{sec:devdataanalysis}
Regular surveys are effective tools for collecting the perceptions of development team members regarding agile practices~\cite{Kropp2018}.
However, they do not allow insights into whether these gathered assessments are rooted in actual project reality, i.e., whether the perception of following a specific practice is traceable in project data.

\subsection{Development Data collection}
Evaluating the development data produced during project work can thus provide another dimension of analysis and takes advantage of the fact that usage of development practices is ``inscribed into software artifacts''~\cite{de2005seeking}.
For every sprint and development team, we collected the produced development artifacts from the course's \emph{git} repository hosted at the collaboration platform GitHub.
GitHub allows programmatically extracting the stored data stored through comprehensive application programming interfaces (APIs)~\cite{GitHubApi}.
The extracted data points included version control system commits as well as tickets from an issue tracker, which was used by teams to manage user stories and the product and sprint backlogs.
User stories included labels (e.g., ``Team A'') and assignments to users which allowed connecting them to developers and teams as well as title and body text, current and previous statuses (e.g., open or closed) and associated timestamps.
Extracted commits contain the committing user ID, the diff (source code line changes), a commit message describing the change and a timestamp when check-in occurred.

\subsection{Measurement Definition}
\label{sec:measurementdefinition}
Based on the background research described in Section~\ref{sec:survey_detail} as well as related research on process metrics~\cite{Rahman:2013:WPM:2486788.2486846} and team evaluation~\cite{Ju2017,Ibrahim2015}, we defined a set of data measurements related to the agile practices mentioned in the survey.
As we had intimate knowledge of development teams, their processes, development tools, and their work environment during the project, we could design measurements specifically for the context, taking advantage of the available data sources.
Therefore, in order to use apply these measurements more generally, adaptations to the different context will be necessary.
However, the development data measurements we employed require no additional documentation overhead for developers, next to the agile development best practices taught in the course, meaning that necessary data is collected in a ``non-intrusive'' manner~\cite{zazworka2010developers}.
Measurements are furthermore designed to be simple for two main reasons:
First, an analytics cold-start problem exists~\cite{Guo2016}, as the project environment is completely new and no previous data is available to draw from, e.g., to base sophisticated prediction models on.
Second, all measurements should be intuitively traceable to the underlying data points, in order to allow comprehension of analysis results by non-expert users and students without detailed explanation.

\subsubsection{Test-Driven Development}
Test-driven Development, as one of the core principles of Extreme Programming, has been around since the late '90s~\cite{Madeyski10}.
As such, a number of approaches and techniques to detect and rate the use of TDD have been developed~\cite{Madeyski10,Buffardi2012,Johnson2007}.
For this study, we follow the approach of Buffardi and Edwards~\cite{Buffardi2012}, which quantifies the usage of TDD based on the measurement of Test Statements per Solution Statement (TSSS).
It relates the number of programming statements in test code to the number of statements in solution code, i.e., code which contains business logic and makes the tests pass.
The programming language employed during our course is \emph{Ruby}.
The Ruby style guide, which students are strongly encouraged to follow, recommends employing a single programming statement per line~\cite{RubyStyleGuide}, therefore lines of code can be used as a proxy for statement amount.
Furthermore, the chosen web development framework \emph{Ruby on Rails}, through the idea of ``convention over configuration''~\cite{Vuksanovic2011}, mandates a strong separation between test, application, configuration and glue code in projects; they all reside in specific directories.
While business logic is contained in an \texttt{app/} folder, test files are included in a separate directory.
Using commit data, this allows calculating the ratio of changed lines (the sum of added, modified and deleted lines) of test and implementation code for every developer in a sprint as a measure of TDD practice application.
\boxbox{0.8}{\textbf{Measurement RTA}: \textbf{R}atio of line changes in \textbf{T}est code to line changes in \textbf{A}pplication code}

\subsubsection{Collective Code Ownership}
Recent research on (collective) code ownership has proposed approaches of assigning ownership of a specific software component to individuals or groups of developers~\cite{greiler15,Bird2011}.
A contributor's proportion of ownership concerning a specific software component is defined as the ratio of commits of the contributor relative to the total number of commits involving that component~\cite{Bird2011}.
However, our focus in this study lies on the individual contributions of development team members to the practice of CCO for a particular given time frame, i.e., a Scrum sprint.
As development iterations are relatively short, ownership of particular software components varies strongly between these time frames.
For a Scrum sprint, therefore, we compute a proxy for an individual developer's contribution to the concept of CCO in the team as the number of unique files that were edited, identified by their commit timestamps.

\boxbox{0.8}{\textbf{Measurement UFE}: Amount of \textbf{U}nique \textbf{F}iles \textbf{E}dited by a developer in a sprint}

\subsubsection{Last-Minute Commits}
Performing required work mostly close to a looming deadline, as is often the case in group collaborations, goes counter to the agile idea of maintaining a ``sustainable pace''~\cite{Lindstrom2004}.
Tutors communicated that students procrastinated~\cite{Ariely2002ProcrastinationDA} as somewhat expected in group work, reporting that it was not unusual to see developers still coding during the sprint review meeting to maybe still fix a bug or finish a user story.
As tutors were available at all Scrum meetings during project work, we were aware of the exact times that team meetings took place and when teams started and finished sprints.
With this information, we extracted commits from the code repository for every developer that were made ``at the last minute'', i.e., 12 hours before the closing of the sprint, usually defined by the sprint review meeting.
We then computed the ratio of a developer's last-minute commits relative to the total number of commits made by the developer in the sprint.
\boxbox{0.8}{\textbf{Measurement LMC}: Percentage of \textbf{L}ast-\textbf{M}inute \textbf{C}ommits within 12 hours before a team's sprint review meeting}

\subsubsection{Average LOC change}
It is in the nature of agile development to require code and software systems to evolve over time and over development iterations due to requirement changes, code optimization or security and reliability fixes.
The term \emph{code churn} has been used to describe a quantization of the amount of code change taking place in a particular software over time.
It can be extracted from the VCS's change history to compute the required line of code (LOC) changes made by a developer to create a new version of the software from the previous version.
These differences form the basis of churn measures.
The more churn there is, i.e., the more files change, the more likely it is that defects will be introduced~\cite{nagappan2005use}.
In this study, we employ a modified version of the ``LinesChanged'' metric used by Shin et al.~\cite{Shin2011} in that we compute the accumulated number of source code lines changed in a sprint by a developer instead of since the creation of a file.
In modern VCS lines can be marked as added, changed, or deleted.
In git, line modifications are recorded as a deletion followed by an insertion of the modified content.
The average LOC change per commit is then computed by summing all insertions and deletions of a developer's commits in a sprint, divided by total amount of that developer's sprint commits.
\boxbox{0.8}{\textbf{Measurement ALC}: \textbf{A}verage \textbf{L}ine of code \textbf{C}hanges per commit by a developer in a sprint}

\subsubsection{Simultaneous User Stories}
Ideally, developers minimize the number of user stories being worked on simultaneously, by keeping as few tasks open as possible~\cite{Sutherland2007}.
This helps reduce error rates and variability~\cite{Middleton2012} and avoids integration problems and missed deadlines at the end of the sprint~\cite{Sutherland2007}.
Gauging whether multiple stories were being worked on simultaneously requires information on the code changes necessary for implementing a story.
As these two types of information are stored in separate systems, i.e., the VCS and the issue tracker, they need to be connected for analysis.
This connection is enabled by the development best practice of linking commits to user stories via their issue tracker number in commit messages, e.g., ``Rename class; Fixes \#123''.
This convention can provide additional information needed to understand the change and can be used to interact with the issue tracker, which parses commit messages for keywords~\cite{GitHubHelp2019}.
We extracted the mentions of user story identifiers from the version control system for every developer and sprint.
The amount of ``interweaving'' ticket mentions, i.e., a series of commits referencing issue A, then issue B, and then A again, was extremely low.
Post hoc interviews revealed that tagging was not viewed as critical by developers and was often forgotten.
However, the problem of too many open tasks per developer naturally only occurs if a developer worked on multiple stories during a sprint; if only one story is mentioned in a developer's commits there is no problem.
As 38\% of developers referenced no or only a single ticket identifier during a sprint (median 1, see Table~\ref{table:dev_data_results}), we focused on this aspect.
The higher the amount of mentioned unique user story identifiers, the higher the chance that too much work was started at the same time.

\boxbox{0.8}{\textbf{Measurement UUS}: Amount of \textbf{U}nique \textbf{U}ser \textbf{S}tory identifiers in commit messages}

\subsubsection{Code Reviews}
Code reviews as part of modern code review procedures using GitHub are facilitated through comments on Pull Requests (PR)~\cite{Bacchelli2013}.
Comments help spread knowledge within teams and can focus on individual lines of code or code fragments as well as overall design decisions.
These techniques are taught to students during the course.
Due to the wide variety of possible feedback that can be given using natural language, measuring the quality of code review comments without human intervention is an ongoing challenge.
However, our data set of extracted PR comments showed two main clusters of developers: those that did not leave any comments in a sprint and those that left two or more.
Developers who did not comment at all are unlikely to have had an impact of the code review process, whereas those developers who commented freely were much more helpful.
We rely on the number of comments to a PR by a developer to measure the intensity of discussion and developer involvement.
\boxbox{0.8}{\textbf{Measurement PRC}: Amount of \textbf{P}ull \textbf{R}equest \textbf{C}omments made by a developer in a sprint}

\subsection{Development Data Analysis Results}
Table~\ref{table:dev_data_results} shows descriptive statistics of the data collected using the six presented development data measures in the examined software engineering course.

\begin{table}[htbp]
    \centering
    \caption{Descriptive statistics of development data measures}
    \label{table:dev_data_results}
    \begin{tabular}{lSrrrrr}
        \toprule
        & RTA & UFE & LMC & ALC & UUS & PRC  \\
        \midrule
        Valid & 121.0 & 168.0 & 124.0 & 124.0 & 168.0 & 168.0  \\
        Missing & 47.0 & 0.0 & 44.0 & 44.0 & 0.0 & 0.0  \\
        Mean & 1.7 & 12.8 & 0.8 & 38.4 & 1.4 & 6.7  \\
        Median & 0.5 & 9.0 & 0.9 & 27.3 & 1.0 & 1.0  \\
        Stdev & 5.7 & 15.4 & 0.3 & 40.8 & 1.5 & 12.0  \\
        Variance & 31.9 & 235.9 & 0.0 & 1661.0 & 2.2 & 142.8  \\
        Skewness & 7.1 & 2.4 & -1.2 & 3.00 & 1.29 & 2.67  \\
        \multicell{Std. Error \\Skewness} & 0.2 & 0.2 & 0.2 & 0.2 & 0.2 & 0.2  \\
        \bottomrule
    \end{tabular} 
\end{table}

\begin{figure}[htb]
    \centering
    \includegraphics[scale=0.45]{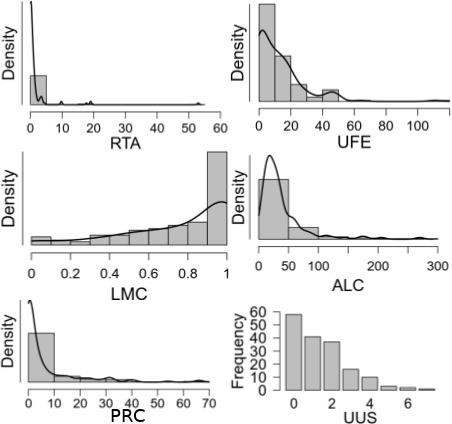}
    \caption{Histograms of development data measurement results.}
    \label{fig:Dist}
\end{figure}

Figure~\ref{fig:Dist} shows histograms including density estimations of the data produced by the six development data measures.
These follow previous software engineering findings, especially in an education context.
The vast majority of developers changed more application code than they modified test code in a sprint (RTA).
Similarly, we identified more developers who edited few unique files in a sprint, between 0 and 20, than developers who edited more than that (UFE).
In accordance with common student teamwork dynamic, many more commits were made to the version control system towards the end of the sprint than were made towards the beginning (LMC).
Developers had followed best practices and had made mostly small and medium-sized commits, with most commits containing up to 50 changed lines of code (ALC).
In line with common issues in software engineering courses was also the fact, that most developers left 10 or fewer comments on Pull Requests helping other students or reviewing code (PRC).
Lastly, too few developers tagged commits with a corresponding issue id, with most developers referencing no more than a single user story per sprint (UUS).

\subsection{Development Data Discussion}
All six development data measures returned actionable results for the data collected in our software engineering course.

\subsubsection{Gaining insights into student teams (RQ~\ref{r5})}
The presented measurements represent consolidated observations of the behavior of development teams concerning selected aspects.
In addition to perceptions gathered through surveys, they allow another perspective based on data.
\boxbox{0.9}{\textbf{Research Question \ref{r5}}:
    How can software development artifacts be used to gain insights into student behavior and the application of agile practices?\\
    \textbf{Answer:} We present six measurements of agile development practice application based on non-intrusively collected project data, see Section~\ref{sec:measurementdefinition}, which can be compared over development iterations.
}
%
\subsubsection{Survey answers and measurements (RQ\ref{r6})}
In order to evaluate, whether the perception of agile values and agile practices are rooted in actual project reality, we repeatedly examined the corresponding relationships by calculating Kendall's Tau and tested its statistical significance by applying Fisher's exact test of independence.
\begin{table}[h]
    \centering
    \caption{Measures of relationship between survey questions Q1-Q9 and development data measures.}
    \label{table:q_to_d_correlations}
        \begin{tabular}{lrrl}
            \toprule
            Relationship &   Kendall’s-$\tau$ & Z & $p$-value  \\
            \midrule
            Q1 - RTA & $-0.361$ & $-5.114$ & $3.1e^{-7}$***  \\
            Q2 - UFE & $-0.022$ & $-0.332$ & $0.740$  \\
            Q5 - LMC & $-0.274$ & $3.738$ & $1.8e^{-4}$***  \\
            Q6 - ALC & $-0.074$ & $-1.059$ & $0.290$  \\
            Q7 - UUS & $-0.052$ & $-0.716$ & $0.474$  \\
            Q8 - PRC & $-0.110$ & $-1.560$ & $0.119$  \\
            \bottomrule
            \multicolumn{4}{l}{* p $<$ .05, ** p $<$ .01, *** p $<$ .001} 
        \end{tabular}
\end{table}
Kendall's Tau and the corresponding $p$-values in Table~\ref{table:q_to_d_correlations} show that two development data measures had a significant relationship to survey claim responses.
There is a significant relationship between Q1 regarding TDD and the RTA measurement ($\tau=-0.362$, $p<.001$).
This indicates that course participants were able to accurately self-assess their success in working in accordance with TDD and that the RTA measurement captured the work that was performed.
Those students who self-assessed that they had followed the test-driven approach during a sprint also had a high ratio of test to application code line changes.
Furthermore, we found a significant relationship between Q5 regarding working shortly before sprint deadline and the LMC measurement ($\tau=-0.274$, $p<.001$).
We conclude that students were able to critically self-assess whether they had worked mostly in a deadline-driven fashion, postponing work until close to the deadline.
This common behavior was captured by the percentage of last-minute commits (LMC measurement).

\boxbox{0.9}{\textbf{Research Question \ref{r6}}:
    What is the relationship between perceptions of agile practices and the development data measurements based on these?\\
    \textbf{Answer:}
    There are two significant relationships: (i) survey answers on TDD usage (Q1) and the RTA measurement, (ii) survey answers on last-minute work (Q5) and the LMC measurement.
}

\section{Conclusion}
\label{sec:conclusion}
In this paper, we investigated software developers' perceptions of agile practices in the context of an undergraduate software engineering course.
We developed a set of survey claims concerning agile practice application to collect these assessments and presented the results of the survey.
We show that the concepts of Collective Code Ownership, usage of the version control system in line with agile ideas, and not working at the last minute, correlated with high self-assessments of agile value application.
Furthermore, we developed a set of six development data measures based on non-intrusively collected software development artifacts, which allow insights into team behaviors.
We show that measurements regarding Test-Driven-Development and last minute work correlate with corresponding self-assessments.
These findings highlight areas where assumptions of project team work were validated as well as those areas where perceptions of agile practices and measurements diverged.
These represent opportunities for further investigation.
In line with related research~\cite{zazworka2010developers}, we consider translating development practices into workable definitions and measures as one of the biggest challenges and opportunities.
By sharing our development data measurements and their background in detail we hope to take another step towards this goal.

\balance

\bibliographystyle{IEEEtran}
\bibliography{main}

\end{document}